# Metamaterial Perfect Absorber Analyzed by a Meta-cavity Model Consisting of Multilayer Metasurfaces

Khagendra Bhattarai[1,5], Sinhara Silva[1,5], Kun Song[2,5], Augustine Urbas[3], Sang Jun Lee[4], Zahyun Ku[3,*] & Jiangfeng Zhou[1,*]

[1]Department of Physics, University of South Florida, Tampa, 33620, USA.
[2]Smart Materials Laboratory, Department of Applied Physics, Northwestern Polytechnical University, Xi'an, 710129, China
[3]Air Force Research Laboratory, Materials Directorate, Wright-Patterson AFB, 45433, USA.
[4]Division of Convergence Technology, Korea Research Institute of Standards and Science, Daejeon, 34113, Korea.
[5]These authors contributed equally to this work.
[*]Correspondence and requests for materials should be addressed to J.Z. (E-mail: jiangfengz@usf.edu), Z.K. (E-mail: zahyun.ku.1.ctr@us.af.mil), or S. L. (E-mail: sjlee@kriss.re.kr,)

## Abstract

We demonstrate that the metamaterial perfect absorber behaves as a meta-cavity bounded between a resonant metasurface and a metallic thin-film reflector. The perfect absorption is achieved by the Fabry-Perot cavity resonance via multiple reflections between the "quasi-open" boundary of resonator and the "closed" boundary of reflector. The characteristic features including angle independence, ultra-thin thickness and strong field localization can be well explained by this model. With this model, metamaterial perfect absorber can be redefined as a meta-cavity exhibiting high Q-factor, strong field enhancement and extremely high photonic density of states, thereby promising novel applications for high performance sensor, infrared photodetector and cavity quantum electrodynamics devices

## Introduction

The metamaterial perfect absorber (MPA), exhibiting nearly unity absorption within a narrow frequency range[1-6], has attracted tremendous interest recently. The MPA possesses characteristic features of angular-independence, high $Q$-factor and strong field localization that have inspired a wide range of applications including electromagnetic (EM) wave absorption[3,7,8], and spectral[5] and spatial[6] modulation of light, selective thermal emission[9], thermal detecting[10], infrared photo detecting[11-14], and refractive index sensing for gas[15] and liquid[16-18] targets. The MPA typically consists of three layers: a metallic resonators layer (e.g., cross-type resonators,[9,19] split-ring resonators[3] or metallic nanoparticles[20], and a highly reflective layer, e.g. metallic film[3,9,19,20] or metallic mesh grid[3], separated by a subwavelength-thick dielectric film (spacer). The mechanism of MPA has been explained through a number of models including impedance matching[3,7,21] and destructive interference[22]. In the impedance matching mechanism, the entire three-layered structure is considered as a thin slab made from an homogeneous medium with frequency-dependent effective permittivity $\epsilon_{\text{eff}}(\omega)$ and effective permeability $\mu_{\text{eff}}(\omega)$. At the perfect absorbing frequency ($\omega_0$), the effective permittivity and the effective permeability reach the same value ($\epsilon_{\text{eff}}(\omega)|_{\omega=\omega_0} = \mu_{\text{eff}}(\omega)|_{\omega=\omega_0}$), so that the impedance (both the real part and the imaginary

part) matches to the free space, $z_{\text{eff}}(\omega)|_{\omega=\omega_0} = z_{\text{eff}}'(\omega_0) + i z_{\text{eff}}''(\omega_0) = 1$. Meanwhile, the effective refractive index $n_{\text{eff}}(\omega)$ at $\omega = \omega_0$ exhibits a large imaginary part, $n_{\text{eff}}''(\omega_0) \gg n_{\text{eff}}'(\omega_0)$, thus incident EM wave enters MPA without any reflection and then rapidly decays to zero inside the MPA. In the destructive interference mechanism, the resonator layer and the metallic film are considered as two decoupled surfaces, and destructive interferences of waves reflected multiple times by two surfaces lead to zero reflection[22,23].

In this work, we use a transfer matrix method to assemble the MPA from three functional layers and obtain the phase and the amplitude conditions for the perfect absorption. Analyzing these conditions, we reinterpret the MPA structure as a meta-cavity bounded between a "quasi-open" boundary of a resonant metasurface and a "closed" boundary of a metallic film. The Fabry-Perot (FP) modes are achieved when the phase condition is satisfied, which leads to a strong resonance with non-perfect absorption (NPA). However, the perfect absorption (PA) only occurs when both the phase and the amplutide conditons are satisfied simultaneously. Although PA has drawn significant attention, NPA are more commonly used in practical applications such as sensors[15-18,24] and detectors[10-14]. Both PA and NPA can be well explained by the meta-cavity model. With an improved retrieval method, we find that the resonant metasurface operates at off-resonance wavelengths, thereby acting as a homogenous thin film with high dielectric constant. The FP model redefines MPA as a meta-cavity with high Q-factor and extremely high photonic density of states (PDOS) where the FP modes are easily tailoered by the geometric parameters of metasurface. Our work can pave the way for novel photonic, optoelectronic and cavity quantum electrodynamic (QED) applicaitions[25].

## RESULTS

**Single-layer Effective Medium Model**     Our perfect absorber consists of three layers: an array of cross-wire resonators, a dielectric spacer ($\varepsilon = 2.28$) and a metal ground plane (MGP) as shown in Fig. 1a. We carry out a 3D full-wave simulation to solve Maxwell's equations and obtains numerical solutions by Computer Simulation Technology Microwave Studio that uses a finite integration technology [26]. The parameters used for simulation are as follows: period of cross-wire array, $p = 2\ \mu m$, length and width of the cross-wire, $l = 1.7\ \mu m$ and $w = 0.4 \mu m$, permittivity of GaAs substrate and a dielectric spacer $\varepsilon_{GaAs} = 11.56$ and $\varepsilon_s = 2.28$, thickness of the dielectric spacer, $t_s = 0.09\ \mu m$, and thickness of cross-wire and ground plane, $t_g = 0.1\ \mu m$. A gold used for cross-wire and MGP is described by a Drude model [27] with plasma frequency, $\omega_p = 1.37 \times 10^{16}\ rad/s$ and collision frequency $\omega_c = 4.08 \times 10^{13}\ rad/s$. We use three different models to describe the perfect absorber: (i) a single-layer effective medium model where the entire structure (Fig. 1a) is considered as a layer of homogenous medium (Fig. 1d) characterized by the effective permittivity ($\epsilon_{\text{eff}}$) and effective permeability ($\mu_{\text{eff}}$); (ii) a three-layer effective medium model where the cross-wire resonator (Fig. 1b) is considered as a homogenous effective film (Fig. 1e) with $\epsilon_{\text{eff}}$ and $\mu_{\text{eff}}$ on top of two layers of real films, the spacer and MGP; (iii) a transmission line model where the multiple-layer optical system (Fig. 1c) is modeled as a two-port network (Fig. 1f) with input impedance $z_{\text{in}}$ and output impedance $z_{\text{out}}$.

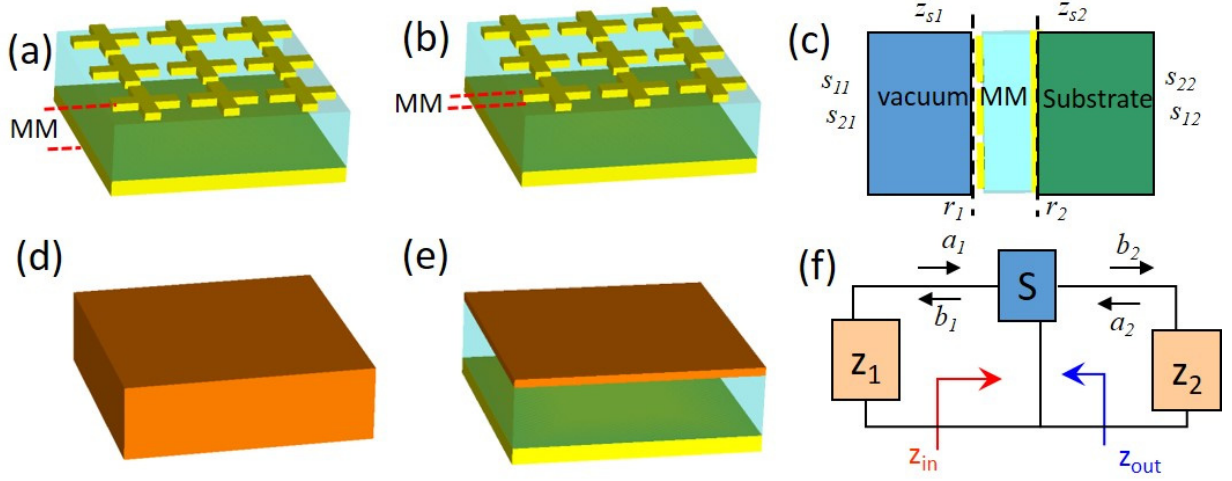

**Figure 1. Effective models for MPA.** Schematic diagram of (**a**) metamaterial perfect absorber (MPA) and its (**d**) equivalent single-layer film characterized by effective $\epsilon_{\text{eff}}$ and $\mu_{\text{eff}}$; (**b**) perfect absorber with cross-wire, spacer and metal ground plate considered as a three-layer structure and the equivalent three-layer film model (**e**) where the cross-wire is considered by a homogenous film with $\epsilon_{\text{eff}}$ and $\mu_{\text{eff}}$ (**c**) View of perfect absorber along wave propagation direction. The MPA is surrounded by air and a dielectric substrate with optical impedances $z_{s1}$ and $z_{s2}$ at two interfaces, respectively. (**f**) A transmission line model uses a two-port network $S$ to describe the MPA, where $z_1$ and $z_2$ are impedance of air and substrate, and $z_{\text{in}}$ and $z_{\text{out}}$ are input and output impedances of $S$, respectively.

Since the MGP prevents the EM wave from propagating through, the perfect absorption is obtained when the reflection reaches zero at certain wavelength as shown in Fig. 2d. Among three models shown in Fig. 1, the single-layer effective medium model is widely used in various MPA work [1,3,7,21], where the electric and the magnetic resonances lead to the impedance matching condition, $z_{\text{eff}} = \sqrt{\mu_{\text{eff}}/\epsilon_{\text{eff}}} = 1$, at the perfect absorption wavelength. Therefore, the incident EM wave propagates through the top surface of effective film in Fig. 1d without any reflection. Meanwhile, the effective film exhibits large imaginary effective refractive index, $n''_{\text{eff}} \gg n'_{\text{eff}}$, so that the EM wave decays exponentially to zero before it exits from the bottom surface. To verify this interpretation, we use effective medium theory [28] to retrieve the effective permittivity ($\epsilon_{\text{eff}}$) and effective permeability ($\mu_{\text{eff}}$) from the complex transmission coefficient $t$ and reflection coefficient $r$. Due to the asymmetric structure of MPA along the propagation direction of incident light, the perfect absorption only occurs for incident illumination from the front-side (cross-wire) and the EM wave incident from the back-side (MGP) will be totally reflected. Strictly speaking, such extremely asymmetric propagation in MPA cannot be described by waves traveling in a simple medium with homogenous permittivity and permeability. However, since we only consider the interaction of MPA with incident light from the front-side (cross-wire), we can use $t$ and $r$ for impinging light from the air-side in order to obtain $\epsilon_{\text{eff}}$ and $\mu_{\text{eff}}$ as shown in Fig. 2a,b, respectively. We find a Lorentzian type resonance of $\epsilon_{\text{eff}}$ at wavelength of 5.13 μm, which can be ascribed to the first-order dipole resonance mode of the cross-wire driven by the electric field of the incident wave. For wavelength $\lambda > 6\mu m$, Re($\epsilon_{\text{eff}}$) is negative. This Drude type response is attributed to the MGP, which exhibits the typical characteristic of permittivity of bulk metal, but with much lower plasma frequency due to diluted average electron density resulting from low volume-ratio of MGP inside the MPA unit cell [29]. The effective permittivity $\mu_{\text{eff}}$ also shows a Lorentzian type resonance at wavelength of 5.94 μm driven by the magnetic field. This magnetic resonance is induced by the anti-parallel currents on the cross-wire and the MGP, and can be qualitatively explained by the method of image theory [7]. The solid lines in Fig. 2c are the real (red color) and the imaginary (blue color) parts of the effective impedance $z_{eff}$, which show strong dispersive response as a function of wavelength.

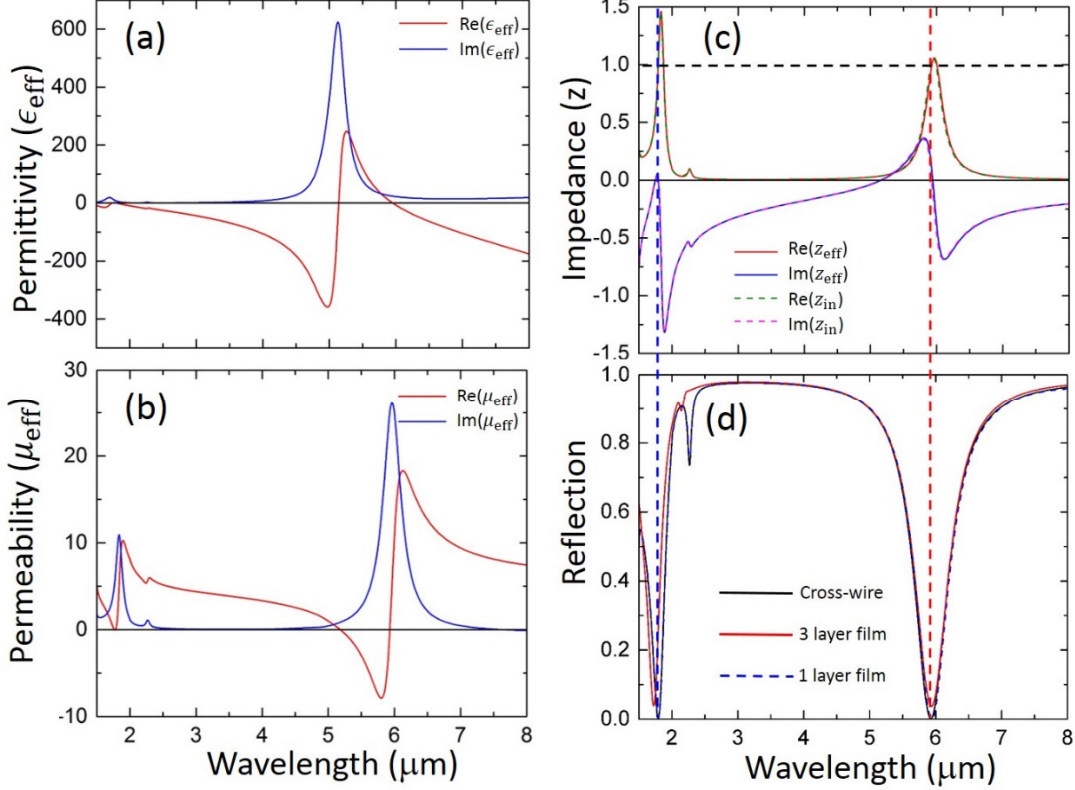

**Figure 2. Effective parameters for single-layer effective medium model.** Real and imaginary parts of (**a**) effective permittivity and (**b**) effective permeability of MPA retrieved by using a single-layer thin film model. (**c**) Real and imaginary parts effective impedance calculated by single-layer thin film model ($z_{eff}$, solid curves) and effective circuit model ($z_{in}$, dash curves). (**d**) Simulated reflection spectra of the actual MPA (black-solid) and structures consisting of a single-layer effective film (blue-dashed) and three-layer effective film (red-solid).

At the perfect absorption wavelength λ = 5.94 μm ($\epsilon_{eff} = 2.432 + 24.175i$ and $\mu_{eff} = 1.003 + 25.517i$), this leads to $z_{eff} = 1.02 + 0.03i$, which matches with the impedance of air ($z_{air} = 1$) very well. The effective refractive index, $n_{eff} = \sqrt{\epsilon_{eff} \cdot \mu_{eff}} = 1.736 + 24.849i$ has a large imaginary part (i.e., $n'' = 24.849$). These calculated effective constitutive parameters very well support the single-layer effective medium model. Specifically, the matched impedance allows the incident wave passing through the top surface of the effective thin film (Fig. 1d) without any reflection. Meanwhile, the intensity of wave is $I = I_0 e^{-2\alpha z}$, where $\alpha = -n''k = -\frac{2\pi n''}{\lambda} = -26.28\ \mu m^{-1}$ (i.e., incident wave decays at a rate of $1/e$ for every 0.019 μm distance traveling inside the MPA). When the incident wave is reflected by MGP and travels a round-trip inside MPA back to the front-surface, its intensity reduces to $I/I_0 = e^{-4\alpha \cdot t_s} = 7.8 \times 10^{-5}$. Thus 99.98% of the wave energy is absorbed inside the MPA. As shown in Fig. 2d, the second absorption peak can be also found at λ = 1.79 μm and the effective impedance $z_{eff} = 1.022 + 0.006i$ is found to be very close air ($z_{air} = 1$) again. Other effective parameters take the following values: $\epsilon_{eff} = 0.3008 + 7.157i$, $\mu_{eff} = 0.378 + 7.8331i$ and $n_{eff} = 0.338 + 7.49i$. To the best of our knowledge, this second absorption peak (λ = 1.79 μm) has not been thoroughly studied in the past. As will be shown later, it results from the second-order dipole resonance of the cross-wire. The effective impedance of the MPA can be obtained by the effective circuit model [30]. Figure 1f shows an effective circuit for the MPA, where the MPA is described as a two-port element $S$ in connection with loads $z_1$ (air) and $z_2$ (substrate). We calculated the input impedance $z_{in}$ based on transmission line theory and plotted it as dashed curves in Fig. 2c. Although $z_{in}$ was calculated with complete different methodology (see details in Methods section), it shows excellent agreement with $z_{eff}$ obtained from single-layer thin film model. Using the retrieved effective permittivity and permeability $\epsilon_{eff}(\omega)$ and $\mu_{eff}(\omega)$, we performed full-wave simulation for the single-layer effective thin film (Fig. 1d). The simulated reflection (blue-dashed curve in Fig. 2d) completely overlaps the reflection from the original absorber structure (black-solid curve) as expected. This further investigation based on the single layer effective medium model can provide fundamental insight for MPA (e.g., reflection and transmission), however this model cannot provide the microscopic mechanism of the absorption, cannot very well explain NPA and can miss some important features such as ultra-thin thickness.

**Three-layer Effective Medium Model.** To better understand the underlying mechanism that the wave travels and decays inside the MPA, we have studied the transmission and reflection of waves at each constituting layers and used a transfer matrix method to obtain the overall reflection/absorption properties of MPA. We considered the cross-wire as a metasurface and described it as a homogenous thin film with effective permittivity $\epsilon_{eff}$, and effective permeability $\mu_{eff}$. The effective parameters were calculted using transmisison and refleciton coefficients of an air/cross-wire/spacer configuration through an improved retrieval method to take account of the asymmetric structure due to different incoming and outgoing medium (air and spacer [31]. As shown in Fig. 3a, the effective permittivity of cross-wire shows a Lorentzian type electric resonance at wavelength of λ$_r$ = 4.75 μm, which is shorter than the electric resonance obtained by the single-layer model (5.94 μm). The resonance at 4.75 μm is purely electric since the effective permeability (Fig. 3b) takes a constant value of Re($\mu_{eff}$) = 0.8 except a typical anti-resonance at $\lambda_r$ due to the periodicity effect [32]. With those values of $\epsilon_{eff}$ and $\mu_{eff}$, we used a thin film of the same thickness to replace the cross-wire in the absorber structure. 3D Full-wave simulation of the three-layer thin film model (Fig. 1e) was carried out. The simulated reflection (red-solid) is plotted in Fig. 2d, which matches very well with simulation of actual MPA struture (black-solid) at the first absorption wavelength (λ = 5.94 μm) and is slightly off at the second absorption wavelength (λ = 1.79 μm). We ascribe the small discrepancy at λ = 1.79 μm to possible inaccurcy of simulaiton which becomes usually notable for short wavelength (second absorption wavelength) relative to long wavelength (first absorption wavelength). A good agreement between the three-layer thin film model and actual MPA proves that the cross-wire layer and the MGP layer are decoupled, i.e. neither the existance of MGP will affect the resoannce of cross-wire nor the resonant current on the cross-wire will affect MGP where the latter only plays a role of a reflecting mirror. Therefore, the transfer matrix method as a common approch to study layered medium can be used to obtain the overall property of MPA from each layers and to investigate the absorption mechanism of EM waves inside the MPA. The overall reflection of the three-layer structure can be calculated by multiplying the transfer matrix of each layer, $M = M_1 \cdot M_2 \cdot M_3$, as given below (more details can be found in our previous works[23,31].

$$r = \frac{r_{12} + \alpha r_{23} e^{2i\beta}}{1 - r_{21} r_{23} e^{2i\beta}} \quad (1)$$

In Eq. 1, $r_{12}$ and $r_{21}$ are the reflection coefficients of the cross-wire from front (air)- and back (spacer)-side, respectively. $r_{23}$ is the reflection coefficient of the MGP. $\beta = n_s \cdot k \cdot t_s$ is the propagating phase in the dielectric layer, where $n_s$, $t_s$ and $k$ are the refractive index, the spacer thickness and the wave vector in free space, respectively. $\alpha = t_{21} t_{12} - r_{21} r_{12}$, where $t_{12}$ and $t_{21}$ are the transmission coefficients through the cross-wire along forward (air/crosswire/spacer) and backward (spacer/cross-wire/air) directions, respectively. Although $\alpha$ is strictly equal to 1 at the interface between two homogenous medium, $\alpha \neq 1$ for the cross-wire around the resonance wavelength. This is caused by the structral asymmetry depending on the propagation direction of incident light ($t_{12} \neq t_{21}$). For perfect absorption (i.e., $r = 0$) according to Eq. 1, we obtain the following conditions for the amplitude and the phase, respectively:

$$|r_{12}| = |\alpha \cdot r_{23}| \quad (2)$$
$$\theta = \phi(r_{12}) - \phi(r_{23}) - \phi(\alpha) - 2\beta = (2n+1)\pi, \quad |n| = 0,1,2,\ldots \quad (3)$$

The coefficients $r_{12}$, $r_{21}$, $t_{21}$ and $t_{12}$ are obtained from simualtion of the air/cross-wire/spacer configuration and $r_{23}$ is obtained from the simulation of the spacer/MGP configuration.

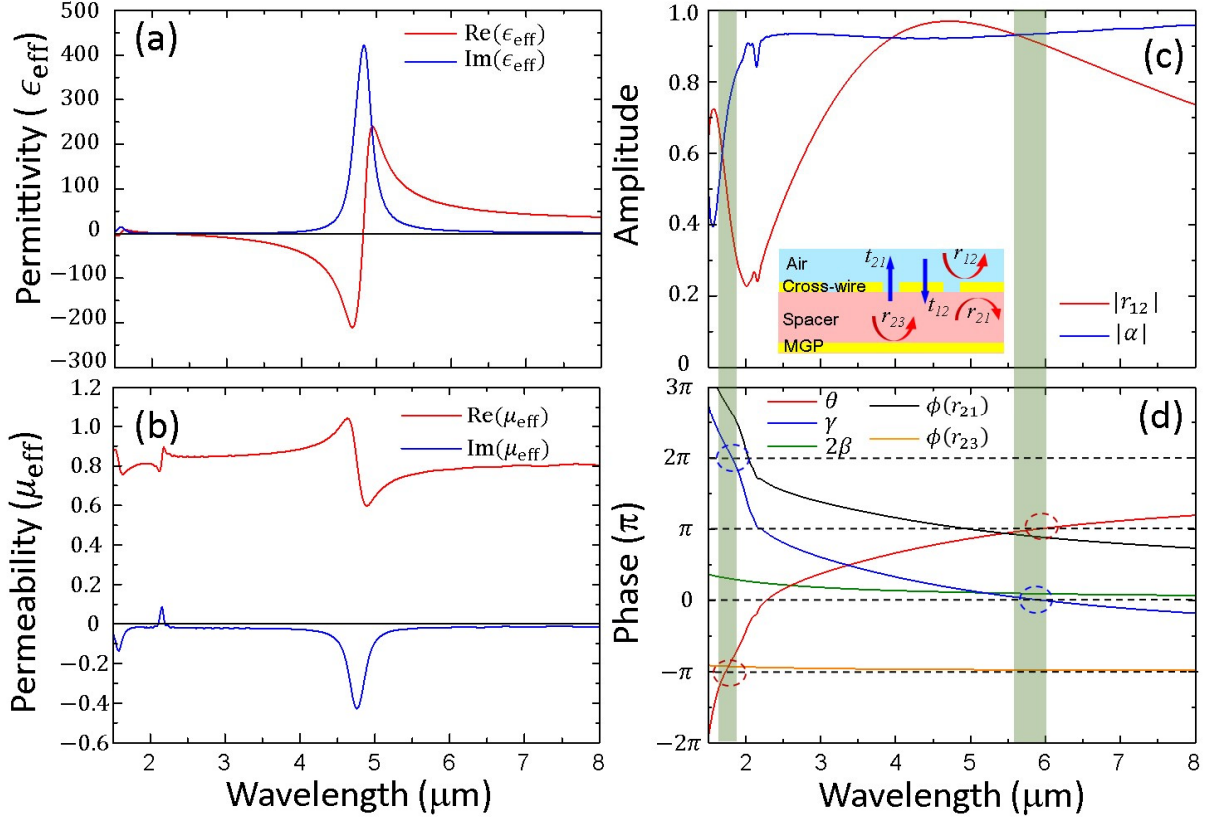

**Figure 3. Effective parameters for three-layer effective medium model and perfection absorption conditions.** Real part (red) and Imaginary part (blue) of (**a**) effective permittivity ($\epsilon_{eff}$) and (**b**) effective permeability ($\mu_{eff}$). (**c**) $|r_{12}|$ (red) and $|\alpha|$ (blue): terms used in the amplitude condition (Eq. 2). Inset depicts various transmission and reflection coefficients of MPA structure. (**d**) $\theta$ (red), $\gamma$ (blue), $2\beta$ (green), $\phi(r_{21})$ (black) and $\phi(r_{23})$ (orange): terms used in the phase condition (Eq. 3,4). Two perfect absorption regions at 5.94 µm and 1.79 µm are highlighted as green, respectively.

Around the perfect absorption wavelength λ = 5.94 µm, both the amplitude condition in Eq. 2 ($|r_{12}| = |\alpha \cdot r_{23}|$) and the phase condition term ($\theta = \pi$) are stasified simultaneously as shown in Fig. 3d. $\theta$ decreases as the wavelength decreases and $\theta$ finally reaches $-\pi$ at λ = 1.79 µm. Meanwhile, the amplitude condition is also satisfied at λ = 1.79 µm, thereby leading to the second absoprtion peak. More carefully examining Fig. 3d, we find that the ampltitude conditon ($|r_{12}| = |\alpha \cdot r_{23}|$ at λ = 5.61 µm) and the phase condtion ($\theta = \pi$ at λ = 5.88 µm) are not satisfied at exact same wavelength. This is due to the fact that the reflection as shown in Fig. 2d does not reach zero ($\min(|r|^2) = R_{min} = 3.68 \times 10^{-4}$ at λ = 5.94 µm). The phase condition in Eq. 3 suggests a Fabry-Perot cavity-like mechanism. The cross-wire acts as a "mirror" that reflects light due to its high effective permittivity and forms a cavity together with the MGP. The incident wave enters the MPA and is reflected multiple times by MGP and cross-wire. The FP cavity resonances establish when the round-trip phase is fullfilled with the following condition:

$$\gamma = \phi(r_{21}) + \phi(r_{23}) + 2\beta = 2m\pi, \quad |m| = 0,1,2,.... \qquad (4)$$

,where $\phi(r_{21})$ and $\phi(r_{23})$ are the phases of reflection coefficients at the interfaces of the spacer/cross-wire and the spacer/MGP (Fig. 3c), respectively. When $\gamma$ equals to $2m\pi$, the waves that travel multiple round-trips between cross-wire and MGP are in phase and hence interfere constructively, thereby leading to resonant cavity modes. As the wavelength decreases, $\gamma$ (blue curve in Fig. 3d) increases monotonically and $\gamma = 0$ and $2\pi$ can be found at λ = 5.94 µm and λ = 1.82 µm, repectively (matching very well with two perfect absorptions wavelengths). Note that the phase condition of Eq. 4 only guarantees the FP cavity modes, resulting in the absorption peaks. Thus, the amplitude condition in Eq. 2 is necessary to be stastisfied at the same wavenegth as well in order to achieve the perfect absorption. Nevertheless, Eqs 3 and 4 well explain the NPA peaks of a varity of resonator/spacer/MGP strcutures that have been widely used in sensing and detecting applications[10-18,24] where the absorption peaks do not necessarily reach 100% (i.e., R = 0). The ultra-thin thickness of MPA can also be explained by the cavity model. As shown in Fig. 3d, at the first perfect absorption wavelenth λ = 5.94 µm where $\gamma = 0$, the phase $\phi(r_{23}) \approx -\pi$ and $\phi(r_{21})$ is slightly less than $\pi$, i.e., $\phi(r_{21}) = \pi - \delta$ with $\delta \ll \pi$. To achieve $\gamma = 0$, it is required $\beta \approx \delta/2$, thus the spacer thickness $t_s$ is found to be $\delta\lambda/4\pi \ll \lambda$. In our simulations, we obtained $\delta = 0.11\pi$ and thus $t_s = 0.0275 \cdot \lambda_s \approx 0.108\ \mu m$, which is close to the actual thickness $t_s = 0.09\ \mu m \approx \lambda_s/44$ with $\lambda_s = \lambda/n_s = 3.94\ \mu m$ being the wavelength inside the spacer medium. This thickness is extremly thin in comparasion to a regular FP cavity with size of $\lambda_s/2$. Note that the key fact that leads to this ultra-small size of cavity is $\phi(r_{21})$ being slightly smaller than $\pi$ due to high effective permittivity of the cross-wire layer.

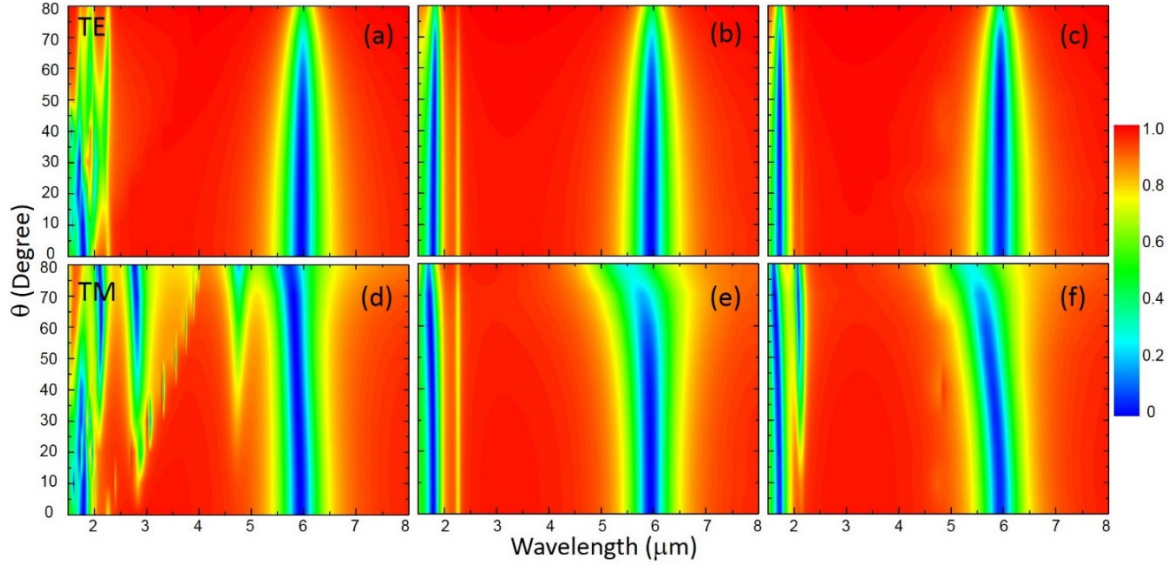

**Figure 4**. **Color plot of angular dependence ($\theta$) of reflection.** Reflection of TE-polarized wave light for (**a**) actual MPA structure, (**b**) single-layer thin film model and (**c**) three-layer thin film model. (**d**)-(**f**) Reflection of TM-polarized wave for corresponding structures.

Another important characteristic feature of MPA is the angular independence of the incident wave. This is very useful in thermal imaging/sensing and energy harvesting applications to receive off-normal incident lights. We performed the simulations of actual MPA structure, single-layer and three-layer thin film model by varying the polar angle $\theta$ of incident wave from 0° to 80°. The reflection spectra are shown in Fig. 4a-c for TE wave and Fig. 4d-f for TM wave, respectively. For TE wave, Fig. 4a shows the wavelength of absorption peak at $\lambda = 5.94$ μm is not shifted as the incident angle increases from 0° to 80°, showing good angular independence. Both the single-layer (Fig. 4b) and the three-layer (Fig. 4c) models are well matched with the actual MPA structure. For TM wave, the reflection of the MPA is blue-shifted as $\theta$ increases by $\Delta_\lambda \equiv \left|\frac{\lambda_{80°}-\lambda_{0°}}{\lambda_{0°}}\right| \sim 4\%$ as shown in Fig. 4d. Both the single-layer and the three-layer models also show the blue-shift with $\Delta_\lambda = 7.4\%$ and 9.1%, respectively, which is slightly higher than actual MPA. The three-layer model can qualitatively explain the angular dependence for TE and TM modes. The cross-wire layer exhibits an electric resonance excited by the electric field of incident wave. For TE mode, the electric field is always parallel to the cross-wire in spite of changing the polar angle $\theta$. Thus, the resonance of cross-wire does not change with incident angle. This leads to the angular independence of absorption for the entire MPA structure. For TM mode, however, the electric field direction changes with $\theta$, resulting in shifting the wavelength of absorption peak.

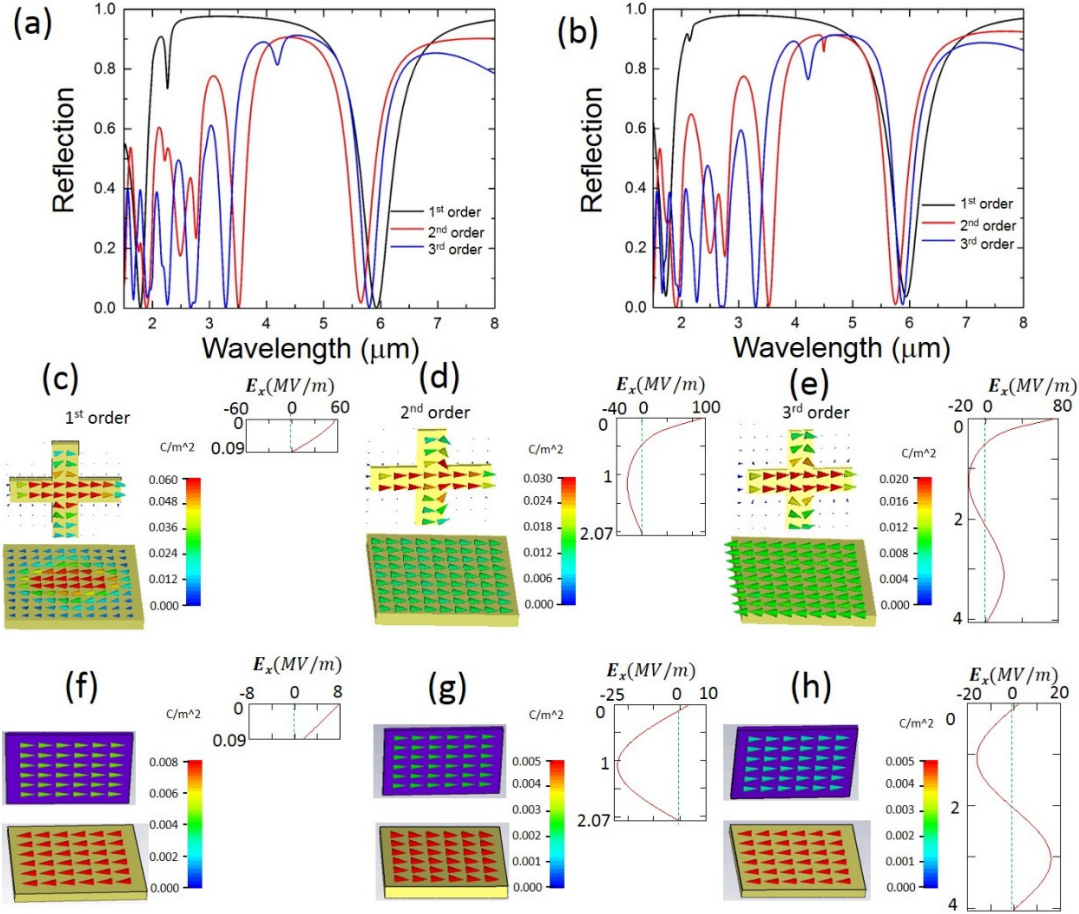

**Figure 5. Reflection, $E_x$ field and current distributions for lowest three orders FP modes.** (**a**) Simulated reflection spectra for actual MPA structures with the spacer thickness of $t_{s1} = 0.09\ \mu m$ (black), $t_{s2} = t_{s1} + \lambda_s/2 = 2.06\ \mu m$ (red) and $t_{s3} = t_{s1} + \lambda_s = 4.03\ \mu m$ (blue). (**b**) The corresponding reflection spectra for three-layer thin film model with spacer thickness $t_{s1} = 0.09\ \mu m$, $t_{s2} = 2.06\ \mu m$ and $t_{s3} = 4.03\ \mu m$. Current density on the cross-wire and MPG for (**c**) 1$^{st}$ order, (**d**) 2$^{nd}$ order and (**e**) 3$^{rd}$ order MPA. Insets placed on right side of (c)-(e) show the x-component of electric field $E_x$ along z-direction (propagation direction). Current density on the meta-film representing cross-wire and the MGP for (**f**) 1$^{st}$ order, (**g**) 2$^{nd}$ order and (**h**) 3$^{rd}$ order. Insets on right side of (f)-(h) display $E_x$ along z-direction for the three-layer thin film structures.

**Higher order FP modes.** Equations 3 and 4 indicate that the phase conditions are satisfied when $\beta$ increases by integer multiples of $2\pi$, thereby leading to a series of FP cavity modes. The increment of $2\pi$ in $\beta$ corresponds to an increase in the thickness of spacer ($t_s$) by $\lambda_s/2$. For the first absorption peak at $\lambda = 5.94\ \mu m$ shown in Fig. 2d, the increment of thickness is calculated as $\lambda_s/2 = 1.97\ \mu m$ and the corresponding n-th order thickness is given by $t_{sn} = t_{s1} + n(\lambda_s/2)$ where $t_{s1} = 0.09\ \mu m$ is the thickness of the original MPA. Figure 5a shows the reflection spectra for the lowest three orders of thicknesses, $t_{s1} = 0.09\ \mu m$, $t_{s2} = 2.06\ \mu m$ and $t_{s3} = 4.03\ \mu m$. The reflection spectra for the thickness $t_{s2} = 2.06\ \mu m$ (red curve) and $t_{s3} = 4.03\ \mu m$ (blue curve) show the absorption peaks at wavelengths of 5.65 μm and 5.81 μm, respectively, which are well matched with the absorption peak at wavelength of 5.94 μm for the original MPA ($t_{s1} = 0.09\ \mu m$). We attribute the minor differences in wavelengths to the slight changes of cross-wire resonances caused by different separations between cross-wire and MGP. Figure 5b shows the reflections of three-layer thin film model using the same thicknesses. The results very well match the reflections of actual MPA (Fig. 5a). We recognize these resonances as the lowest three orders of FP cavity modes, which is confirmed by observing the x-component of electric field $E_x$ inside the MPA structure. The right-side figures in Fig. 5c-e show the real part of $E_x$ along z-direction inside the spacer at the wavelength of absorption peaks, where $E_x$ is evaluated along a straight line in z-direction below the middle ($x = 1\ \mu m, y = 0$) of cross-wire (from z = 0, the bottom of cross-wire to z = $t_{sn}$, the spacer thickness). For the first-order thickness $t_{s1} = 0.09\ \mu m$, Re ($E_x$) decreases monotonically from positive value to zero as z increases from 0 (at the cross-wire/spacer interface) to 0.09 $\mu m$ (at the spacer/MGP interface). Such $E_x$ distribution results from the boundary conditions. At the spacer/MGP interface, the high conductivity of MPG forms a "closed" boundary where the tangential electric field is forced to be zero ($E_{t,x} = 0$). At the cross-wire/spacer interface, the high effective permittivity ($\epsilon_{eff} = 64.72 \cdot (1 + 0.11i)$ at λ = 5.94 μm) of cross-wire provides a "quasi-open" boundary which allows non-zero tangential electric field ($E_{t,x} \neq 0$). As the thickness of spacer increases to $t_{s2} = 2.06\ \mu m$, a half-period of sine function curve is added to $E_x$ field for the first order thickness ($E_x$ shown in right side of Fig. 5c) as shown in Fig. 5d, which results in forming the second-order cavity mode. Similarly, $E_x$ for $t_{s3} = 4.03\ \mu m$ (Fig. 5e) shows the fundamental mode plus a full-period of sine curve, leading to the third-order cavity mode. The three-layer thin film model gives the same $E_x$ profiles as shown in Fig. 5f-h. All the $E_x$ fields for actual MPA (Fig. 5c-e) and for effective thin film model (Fig. 5f-h) exhibit constant phases (not shown here) along z-direction, thereby indicating standing waves as we expected according to the abovementioned FP cavity analysis. The left-side figures in Fig. 5c-e show the electric current distributions on the cross-wire and MGP for the lowest three orders of FP modes. In all three figures, the currents on the cross-wires show a half-wavelength cosine function profile along the horizontal direction with maximum in the center and zero at ends. This reveals a first-order dipole resonance mode of the cross-wires and matches with the Lorentzian resonance of $\varepsilon_{eff}$ shown in Fig. 3a. Figures 5c shows the current on the MGP flowing toward the opposite direction with the maximum current in the center region. The anti-parallel currents on the cross-wire and MPG in Fig. 5c induce the strong magnetic moment that is responsible for the magnetic resonance for single-layer thin film model shown in Fig. 2b. We also observe the same anti-parallel currents in the simulation for the three-layer thin film model as shown in Fig. 5f. Such current distributions can be well explained by the $E_x$ profile shown in Fig. 5f. The tangential electric field $E_{t,x}$ induces the electric current, $J_x = i\omega\epsilon E_{t,x}$, where the $\epsilon$ for cross-wire and MGP are given by $\epsilon_{eff}$ and $\epsilon_{MGP}$ (using the Drude model[27]), respectively. At wavelength of λ = 5.94 μm, the real parts of $\epsilon$ for cross-wire and MGP have the opposite signs, i.e., Re($\epsilon_{eff}$) > 0 and Re($\epsilon_{MGP}$) < 0. For the first-order cavity shown in Fig. 5c,f, $E_x$ in metafilm and MGP are in phase, and thereby result in anti-parallel currents. For the higher order cavities (e.g. Fig. 5g,h for the second-order and third-order cavities, respectively), when $t_s$ increases by $\lambda_s/2$, the phase of $E_x$ in MGP increases by $\pi$ and the current in MPG changes to the opposite direction.

We expect the FP cavity model for MPA to lead to novel applications. As one example, the MPA provides much stronger local electromagnetic fields enhancement than typical plasmonic resonators due to its large Q-factor. Such strong enhancement is desired for in sensors and detectors at infrared and THz regimes[11,18,33]. However, the region between resonator and MGP in the MPA, where the fields are mostly enhanced, are too thin to fit most sensing and detector devices. With the cavity model, we can design the higher order cavities that can fit in the required thickness of practical devices. For instance, infrared focal plane array device working at 6 μm wavelength[33], is composed of active layer (20 stacks of InGaAs/AlGaAs quantum wells) and top/bottom contact layers (n-doped GaAs). The total thickness is about 2 μm which is much larger than the first-order MPA thickness of 0.09 μm, but is very close to the second-order thickness of 2.06 μm. Furthermore, this thickness is tunable by varying the size of the resonator (cross-wire) to fit the requirement of practice devices. Another potential application is to use high photonic density of states (PDOS) for cavity quantum electrodynamic (QED) applications[25,34,35]. PDOS is inversely propotioinal to the volume of a cavity. The first-order cavity of MPA has thickness of $\sim\lambda/44$, much smaller than typical FP cavity with the size of $\lambda/2$. This size is also much smaller than typical plasmonic resonators [25] which EM fields normally extend to distances of a few of wavelengths. Our calculation shows the MPA cavity can improve the PDOS by ~80 times as compared to plasmonic resontors. Such high PDOS promises great potential in cavity QED applications[25,34,35].

## Discussion

In summary, we developed a multi-layered metasurface model for MPA. With a transfer matrix analysis, we found a FP cavity mechanism that models the MPA as a cavity bounded between the "qusi-open" boundary of resonator and the "closed" boundary of MGP. The FP cavity model well explains the characterstic features including unltra-thin thickness and angular indepencence. We also found higher order cavity modes when the thickness of MPA (spacer) increases by multiples of half-wavelength. The strong field enahncements of cavity resoannces and high PDOS promoise novel applications in sensing, detecting and cavity QED devices.

## Methods

The effective impedance $z_{in}$ of the circuit model is calculated from the reflection coefficients $\Gamma_{in}$ and $\Gamma_{out}$ at the the input and output termial of $S$, using the following equations [30]:

$$\Gamma_{in} = S_{11} + \frac{S_{12}\Gamma_{out}S_{21}}{1-r_2 S_{22}} \quad (5)$$

$$\Gamma_{out} = S_{22} + \frac{S_{12}\Gamma_{in}S_{21}}{1-r_1 S_{11}} \quad (6)$$

where $S_{11}$, $S_{12}$, $S_{21}$ and $S_{22}$ are the S-parameters, which can be obtained from simulations of entire MPA structure. The coupled equations (Eqs 5 and 6) can be solved analytically:

$$\Gamma_{in} = -\frac{1}{2(s_{11}+s_{12}s_{21}s_{22}-s_{11}s_{22}^2)}(s_{12}^2 s_{21}^2 - 2s_{11}s_{12}s_{21}s_{22} + (1+s_{11}^2)(-1+s_{22}^2) + \sqrt{-4(s_{11}s_{12}s_{21} + s_{22} - s_{11}^2 s_{22})^2 + (1 - s_{12}^2 s_{21}^2 + 2s_{11}s_{12}s_{21}s_{22} + s_{22}^2 - s_{11}^2(1+s_{22}^2)^2)^2}) \quad (7)$$

$$\Gamma_{out} = \frac{1}{(-2s_{11}s_{12}s_{21}+2(-1+s_{11}^2)s_{22})}(s_{12}^2 s_{21}^2 - 2s_{11}s_{12}s_{21}s_{22} + (-1+s_{11}^2)(1+s_{22}^2) + \sqrt{-4(s_{11}s_{12}s_{21} + s_{22} - s_{11}^2 s_{22})^2 + (1 - s_{12}^2 s_{21}^2 + 2s_{11}s_{12}s_{21}s_{22} + s_{22}^2 - s_{11}^2(1+s_{22}^2)^2)^2}) \quad (8)$$

We then solved the input and output impedances from $\Gamma_{in}$ and $\Gamma_{out}$:

$$z_{in} = \frac{1+\Gamma_{in}}{1-\Gamma_{in}} \quad (9)$$

$$z_{out} = \frac{1+\Gamma_{out}}{1-\Gamma_{out}} \quad (10)$$

For MPA, we only consider the input impedance $z_{in}$ as shown in Fig. 2c for the incident wave from the front side.

## Acknowledgments

The USF portion of this work was supported by the Alfred P. Sloan Research Fellow grant FG-BR2013-123 and by KRISS grant GP2016-034. The KRISS portion of this work was supported by the KRISS grant GP2016-034 and the AOARD grant FA2386-14-1-4094 funded by the U.S. government (AFOSR/AOARD).

## Author Contributions

K.B. performed numerical simulations. K.B., S.S., S.K., Z.K. and J.Z. developed the multiple layer model and retrieval method, and performed analytical analysis of the MPA mechanism. J.Z., Z.K. and K.B. wrote the manuscript. S.J.L., A.U., Z.K. and J.Z. supervised the work.

## Additional Information

**Competing financial interests:** The authors declare no competing financial interests.